\documentclass[letterpaper,twocolumn,10pt]{article}
\usepackage{usenix2019_v3}

\usepackage{tikz}
\usetikzlibrary{shapes, arrows, positioning}
\usepackage{varwidth}
\usepackage{amsmath}
\usepackage{listings}
\usepackage{hyperref}
\usepackage[output-decimal-marker={,},exponent-product=\cdot]{siunitx}

\usepackage{filecontents}

\usepackage{algorithm}
\usepackage[noend]{algpseudocode}

\usepackage{pgfplots} 
\pgfplotsset{compat=newest}

\newcommand{\PP}{\emph{Prime+Probe}}
\newcommand{\FR}{\emph{Flush+Reload}}
\newcommand{\PA}{\emph{Prime+Abort}}
\newcommand{\ET}{\emph{Evict+Time}}
\newcommand{\Sbox}{\texttt{S-Box}}

\newcommand{\attackName}{CacheSniper}
\newcommand{\attack}{\textsc{\attackName}} 

\title{\attack : Accurate timing control of cache evictions}

\author{
{\rm Samira Briongos}\\
NEC Laboratories Europe
\and
{\rm Ida Bruhns}\\
Universität zu Lübeck
\and 
{\rm Pedro Malag\'on}\\
ETSIT-LSI-DIE Universidad Polit\'ecnica de Madrid
\and
{\rm Thomas Eisenbarth}\\
Universität zu Lübeck
\and
{\rm José M. Moya} \\
ETSIT-LSI-DIE Universidad Polit\'ecnica de Madrid
}

\begin{document}
\maketitle

%
\begin{abstract}
Microarchitectural side channel attacks have been very prominent in security research over the last few years. Caches have been an outstanding covert channel, as they provide high resolution and generic cross-core leakage even with simple user-mode code execution privileges. To prevent these generic cross-core attacks, all major cryptographic libraries now provide countermeasures to hinder key extraction via cross-core cache attacks, for instance avoiding secret dependent access patterns and prefetching data.
In this paper, we show that implementations protected by `good-enough' countermeasures aimed at preventing simple cache attacks are still vulnerable. 
We present a novel attack that uses a special timing technique to determine when an encryption has started and then evict the data precisely at the desired instant. This new attack does not require special privileges nor explicit synchronization between the attacker and the victim. One key improvement of our attack is a method to evict data from the cache with a single memory access and in absence of shared memory by leveraging the transient capabilities of TSX and relying on the recently reverse-engineered L3 replacement policy. We demonstrate the efficiency by performing an asynchronous last level cache attack to extract an RSA key from the latest wolfSSL library, which has been especially adapted to avoid leaky access patterns, and by extracting an AES key from the \Sbox\ implementation included in OpenSSL bypassing the per round prefetch intended as a protection against cache attacks. 
\end{abstract}

\section{Introduction}

Modern CPUs are highly optimized to achieve the maximum performance and efficiency for existing manufacturing technologies while ensuring logical correctness and isolation between processes. But since any running software interacts with the microarchitectural elements of the processor and  changes their state, shared hardware resources are influenced by all processes running on a machine. An attacker can leverage these shared resources to recover secret information from a victim software by exploiting the microarchitectural state~\cite{Kocher2018spectre,217478,saad_spoiler,Schwarz2019ZombieLoad,canella2019fallout,vanbulck2018foreshadow,smoth2019,ridl}.

Among the microarchitectural elements utilized in these attacks, the cache memory has probably been the most exploited hardware resource. Traditionally, cache side channel attacks have focused on cryptographic implementations. For instance, cache attacks have successfully retrieved ECDSA~\cite{ecdsayarom}, RSA~\cite{lastlevel,inci2015seriously,Yarom2017} and AES~\cite{irazoqui14wait,app9050944,sca} keys, even breaking the isolation between virtual machines (VMs)~\cite{sca,Ristenpart09,Inci2016}. As these attacks do not require any special privileges to succeed, side channel attacks pose a great threat and keep gaining attention.

As a direct response to the threat imposed by microarchitectural attacks, many different countermeasures were put in place. As described by Ge et al.~\cite{ge2018aSurvey}, preemptive countermeasures try to help in the hard task of designing leakage free code~\cite{Wichelmann2018,182946,Brotzman2018CaSymCA,Irazoqui:2018:MPM:3176258.3176316}, hardware countermeasures either design or take advantage of hardware features to avoid the leakage and detection based countermeasures try to determine whether there is an attack going on~\cite{ChiappettaEtAl2016,Briongos2016,Briongos_2018,gulmezoglu2019fortuneteller,Zhang2016}. In this paper, we will focus on the preemptive countermeasures of \emph{side channel resistant code} and \emph{prefetching}. 

Prefetching is used as a strategy to improve performance as well as preventing attackers from observing the cache state in cache attacks~\cite{irazoqui14wait}. For example, a prefetching strategy was implemented in OpenSSL 1.0.0a and beyond: The \Sbox\ implementation prefetches the \Sbox\ before executing each round~\cite{Ashokkumar2019sbox, Polyakov2006aggressively}. If the attacker flushes a line holding data from the \Sbox\ before or during the execution of any intermediate round, and waits till the encryption has been performed, she will only observe cache hits. Consequently, she would not be able to distinguish whether those access are due to the actual utilization of the line or to the prefetch stage~\cite{Ashokkumar2019sbox}. 

Several recent works have shown that in scenarios where the attacker controls the OS, even smallest leakages can be exploited. Many recent prominent works have either exploited the intra-core resource sharing of simultaneous multithreading (SMT)~\cite{gras2020absynthe,bhattacharyya2019smotherspectre,ridl,vanbulck2020lvi} or the ability of interrupted execution of SGX~\cite{van2017sgx,moghimi2020medusa,moghimi2020copycat,aldaya2020one}. These attacks are so powerful that several cryptographic libraries, including OpenSSL, now ignore them: the cost of protecting against them is exuberant and arguably not justified in scenarios where attacker control of the OS is out of scope (or simply worse than the attacks). Thus, they make do with the imperfect but allegedly sufficient countermeasures of prefetching and leakage minimization described above.

In this work, we show that even a classic user-lever cache adversary can overcome these countermeasures. At a very high level, the attacker has an offline \emph{stakeout} phase where she determines the time elapsed between the start of the target algorithm and the use of the target function or data. In the online phase, she then detects the start of the target algorithm by the victim, waits for the time previously determined, and evicts the target data from the cache. The attacker can then collect valuable data to extract the key. Due to the high precision gained from the exact timing of the eviction, very small windows of opportunity can be leveraged. We demonstrate the attack on an OpenSSL AES \Sbox implementation and a square and multiply implementation from wolfSSL. Both are considered protected from side channel attacks, but both have vulnerabilities that can be exploited by \attack.  

\subsubsection*{Contributions}
In summary, this paper presents the following contributions. 
\begin{itemize}
\item[-] We present \attack , a methodology to evict data from the cache at the desired instant, even in the absence of shared memory, taking advantage of the replacement policy. 
\item[-] We reduce the number of samples required to get a cryptographic key by precisely selecting the instant of time at which we desire to interrupt a victim process.
\item[-] We demonstrate that prefetching data to the cache is not effective against cache attacks, if there is a window of time during the execution of a sensitive part of code, at which private data is not accessed uniformly.
\item[-] We show in realistic experimental setups the feasibility of side channel last-level cache attacks against the AES \Sbox\ implementation that includes prefetching as countermeasure, and against an RSA implementation protected by a side channel resistant implementation.
\end{itemize}

\subsubsection*{Disclosure and Code Publication}
We demonstrate our method by attacking two real world cryptographic implementations, an AES \Sbox implementation by OpenSSL (version 1.0.2k) and an RSA square-and-multiply implementation by wolfSSL (version 4.4.0). We responsibly disclosed to both libraries on June 22nd and 23rd. The communication with the OpenSSL team is still ongoing. No CVE was issued since \attack falls outside their threat model. Additionally, the \Sbox implementation was removed entirely from the latest version of OpenSSL. wolfSSL immediately issued CVE-2020-15309 and proposed a fix for the vulnerability, which we checked and acknowledged.

\section{Background}
This section introduces some basic concepts on cache memory, cache attacks and transactional execution that are of key importance in the efficiency of the proposed cache attack.

\subsection{Cache architecture}

Caches are small memories located between the processor and the main memory, specially designed to reduce the gap between processor and memory throughput. Modern processors include cache memories that are hierarchically organized; low level caches (L1 and L2) are core private, smaller and closer to the processor (with reduced latency), whereas the last level cache (LLC or L3) is bigger and shared among all the cores. Intel processors have traditionally included L3 inclusive caches, in order to simplify the implementation of cache coherence: all the data which is present in the private low-level caches has to be in the shared L3 cache.

Most modern processors include $w$-way set-associative caches; a trade-off between directly mapped caches, usually with high cache miss rates, and fully associative caches, with a very complex logic. The cache is organized into multiple sets ($s$), each of them containing $w$ lines of usually 64 bytes of data. The set in which each line is placed is derived from its address (directly mapped). The address bits are divided into offset (lowest-order bits used to locate data within a line), index ($\log_2(S)$ consecutive bits starting from the offset bits that address the set) and tag (remaining bits which identify if the data is cached). Many caches additionally contain slices. The entries are mapped to the slices by a function $f$, which usually depends on some fixed bits of the data. Slice selection mechanisms are usually not public, and effort has gone into reverse engineering them. The do however not have an impact on \attack .

\subsection{Replacement policies}

At some point during the use of a set-associative cache, data will need to be loaded into the cache after a cache miss, but the corresponding cache set is full. At this point, an algorithm called the \emph{replacement policy} decides which of the currently cached lines is evicted and replaced with the line accessed. As it is crucial for maximizing the hit ratio and achieving good performance, manufacturers do not publish the implementation details of the policy. Each cache level has it's own replacement policy. Based on the observations of the evictions in the LLC, most modern processors implement pseudo-LRU (Last Recently Used) policies.  Regarding Intel, the official name of the LLC replacement policy is ``The Quad-Age LRU''~\cite{7476478}.

There have been several efforts to reverse engineer the replacement policy of Intel processors in order to estimate/measure the number of misses, without explaining which concrete elements in the cache would be evicted in the event of a miss~\cite{reversePol1, reversePol, pagewalk-coherence}. Later studies~\cite{gruss2016rowhammer, vila2018theory} have focused on eviction strategies related to maximize the number of evictions in order to improve memory attacks. Recently, the replacement policy of all the cache levels of modern Intel processors have been reverse engineered and published\cite{Briongos2019,Abel19,vila2019cachequery}.

\subsection{Transactional memory and Intel TSX}

Intel TSX is an instruction set extension for x86 that supports Transactional memory. Transactional memory enables optimistic execution of the transactional code regions specified by the programmer. The processor executes the specified sections assuming that there is no conflict with other threads or CPU cores, which might access or modify the same data. Transactional memory reduces the need of mutual exclusion mechanisms, using a local version of data and registering a hardware-based callback mechanism in case a conflict with other threads is detected. If the execution ends successfully, the processor commits all the changes as if they had occurred instantaneously, becoming visible to the remaining processes. Otherwise, the transaction is cancelled, all memory changes are discarded and a callback function is called. This process is known as an \texttt{Abort}, and the callback is known as an \texttt{Abort handler}. There are various reasons why a transaction may abort in Intel TSX, but we particularly focus on the cache related ones. Namely, a transaction aborts if data from its ``write set'' is evicted from the L1 cache or if data from its ``read set'' is evicted from the L3 cache ~\cite{transact15,disselkoen17}. AMD provides similar transaction mechanisms and ARM is introducing it.

\subsection{Related work}
Cache memory was first mentioned as a covert channel in 1992~\cite{Hu92}. Since then, many different techniques have been developed: Kelsey entertained the idea of attacks based on cache hit ratios, Osvik et al. proposed the widely known \ET\ and \PP\ attacks, revealing the cache sets accessed by the victim, and Gullasch and Yarom and Falkner both developed a powerful attack that exploits shared memory, which they later named \FR\  ~\cite{DBLP:journals/jcs/KelseySWH00, OsvikEtAl2006, GullaschEtAl2011, YaromEtAl2014}. From these attacks, \FR\ and \PP\ are widely used due to their high resolution. This work focuses on \FR\ and \PA , a derivative of \PP\ that leverages transactional aborts to detect cache evictions caused by the victim process~\cite{disselkoen17}.

The \FR\ technique requires shared memory, which means the victim and attacker use the same data during their respective execution. This can be met by them using the same library, which is often the case for libraries shipped with the operating system. In addition, a \texttt{clflush} instruction is needed on the target processor, which is not the case in many scenarios, e.g. when attacking from JavaScript~\cite{genkin2018drive}. The attacker uses this instruction to \textit{flush} the desired lines from the cache, making sure the victim process needs to load them from memory to use them. She then reloads the data, measuring the time this takes. If the victim process used the data, the reload time observed by the attacker will be short. This attack is easy to implement and provides precise information about the data the victim process uses at cache-line granularity. 

The \FR\ attack has been used in many ways since it was introduced. Gullasch et~al.~retrieved an AES key and Yarom et~al.~ demonstrated that one trace is enough to retrieve an RSA key, and attacked ECDSA ~\cite{YaromEtAl2014, ecdsayarom, GullaschEtAl2011}. \FR\ is applicable to launch cross-VM-attacks, used for attacking AES, retrieve keystrokes or profile the victim~\cite{irazoqui14wait, Ristenpart09, DBLP:journals/popets/IrazoquiIES15,GrussEtAl2015}. 

It is vital for \FR\ that the attacker and the victim share memory. The \PP\ attack still works in virtual environments \cite{inci2015seriously,lastlevel}. An attacker can target the LLC cache, particularly one set of the LLC, and it will still able to extract sensitive information. Since \PP\ does not require special OS features, it can be applied on virtually any system. As a preparation step for a \PP\ attack, the attacker needs to construct an eviction set (a group of $w$ different addresses that map to one specific set in $w$-way set-associative caches). Constructing eviction sets and dealing with missing address information as well as slice selection mechanisms has been discussed extensively in the literature~\cite{lastlevel, sca, Inci2016, vila2018theory}. 

One of the drawbacks of both the \FR\ and \PP\ technique is the necessity for precise timers to detect whether the victim accessed the memory. The timer-less attack \PA\ exploits Intel's implementation of Hardware Transactional Memory TSX: it starts a transaction to fill the the targeted cache set (prime), then waits if it receives an abort because the victim has accessed this set~\cite{disselkoen17}. However it does obviously require Intel TSX, which is not available on all machines. 

The OpenSSL AES implementation protected by the prefetch was attacked by Ashokkumar et~al. in 2018. They used a chosen plaintext approach targeting the first and second round of the AES in OpenSSL~\cite{Ashokkumar2019sbox}. Cohney et~al. attack the last round of the AES, but in a very different setting. They target a deeper layer implementation of T-Table AES used only to encrypt seeds for the pseudo random number generator in AES~\cite{Cohney2019Pseudorandom}. While failing to protect these inner implementations the same way as the public ones is a notable oversight, it is also one of the main difference to our target: We attack an implementation that is explicitly protected from cache side channel attacks. The second difference is that as many other side channel attacks, this one uses SGX enclaves in step mode to enhance their timing granularity. Moghimi et~al.~\cite{DanielCachezoom} also target SGX assuming a malicious or compromised operating system. The scenario allows them to use the L1 cache as source of information. This way they retrieve various samples per round and manage to distinguish the prefetching stages from the normal operations of the round for both  T-Table and an \Sbox\ implementations of AES.

\attack\ does not require SGX, nor the frequent interrupts of the victim given by the powerful adversarial scenario controlling the OS. It works across cores and does not need special privileges. It only requires either shared memory with the vicitim (e.g. vial libraries) or TSX. 
Thus, \attack\ is a general attack technique that can be used in many settings. We chose to demonstrate it on an AES and and RSA implementation, but it can be applied in many scenarios to circumvent cache attack countermeasures. 

\section{Target algorithms}

The different cryptographic algorithms that have been shown to be vulnerable to cache attacks have in common that they perform secret dependant accesses to memory. An attacker can observe these accesses through the cache and then retrieve the aforementioned secrets. 

For this reason a common countermeasure with less impact on performance than flushing the data from the caches is to prefetch the data in the cache to prevent the attacker from distinguishing data that have been used for the cryptographic operation from data that have not.

We focus on two common algorithms (AES and RSA) that have been targeted multiple times, in particular we focus on two implementations that have been theoretically protected against these attacks. In the following sections we describe their implementations and explain the countermeasures they implement.

\subsection{AES \Sbox\ implementation in OpenSSL}
\label{subsec:aes_background}

AES~\cite{DaemenR02} is a commonly used symmetric block cipher that operates with data in blocks of 16-bytes. It consists of different operations (\emph{AddRoundKey}, \emph{SubBytes}, \emph{ShiftRows} and \emph{MixColumns}) that are repeated each round. The \Sbox\ is the table that holds the data for the \emph{SubBytes} operation, concretely it holds 256 byte values.

The \Sbox\ software implementation of AES replaced the previously used T-Table implementation, which used four tables (T-Table) with pre-computed values of the \emph{SubBytes}, \emph{ShiftRows} and \emph{MixColumns} operations. That is, it transformed the aforementioned operations into look up operations in order to improve the performance of the encryption and decryption processes. The accesses to the tables are key dependant and not all of them are used during the encryption process. This fact has been exploited multiple times to recover the secret keys \cite{irazoqui14wait,app9050944,sca}.

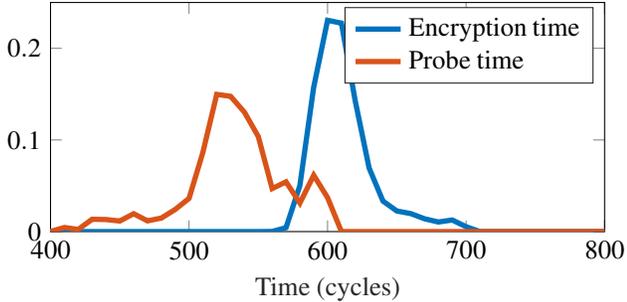
\begin{figure}
	\centering
%
%
\definecolor{mycolor1}{rgb}{0.00000,0.44700,0.74100}%
\definecolor{mycolor2}{rgb}{0.85000,0.32500,0.09800}%
\begin{tikzpicture}

\begin{axis}[%
width=2.9in,
height=1.2in,
at={(1.372in,0.564in)},
scale only axis,
xmin=400,
xmax=800,
xlabel style={font=\color{white!15!black}},
xlabel={Time (cycles)},
ymin=0,
ymax=0.25,
axis background/.style={fill=white},
legend style={legend cell align=left, align=left, draw=white!15!black}
]
\addplot [color=mycolor1, line width=2.0pt]
  table[row sep=crcr]{%
400	0\\
410	0\\
420	0\\
430	0\\
440	0\\
450	0\\
460	0\\
470	0\\
480	0\\
490	0\\
500	0\\
510	0\\
520	0\\
530	0\\
540	0\\
550	0\\
560	0.000125928724342022\\
570	0.00427108256726693\\
580	0.0502560550728288\\
590	0.157694245057298\\
600	0.230575494270243\\
610	0.227500734584225\\
620	0.14238341098938\\
630	0.0693552449313688\\
640	0.0332661713470176\\
650	0.022478277295051\\
660	0.0196553750577173\\
670	0.0139990765226882\\
680	0.0104520841203879\\
690	0.0126348486756496\\
700	0.00535197078453595\\
710	0\\
720	0\\
730	0\\
740	0\\
750	0\\
760	0\\
770	0\\
780	0\\
790	0\\
800	0\\
};
\addlegendentry{Encryption time}

\addplot [color=mycolor2, line width=2.0pt]
  table[row sep=crcr]{%
400	0.000112306047456043\\
410	0.00445181172115756\\
420	0.00234944251278043\\
430	0.0134318032757428\\
440	0.0131532842780518\\
450	0.0115270927108883\\
460	0.0194963298383691\\
470	0.0116843211773268\\
480	0.0147480301519276\\
490	0.0241592769287441\\
500	0.0361939929741337\\
510	0.0865115944763394\\
520	0.149712945742702\\
530	0.147619561018122\\
540	0.130261538323316\\
550	0.103645005076233\\
560	0.0469304511109314\\
570	0.0542842510983531\\
580	0.0311851432575941\\
590	0.0614718381355399\\
600	0.0370699801442908\\
610	0\\
620	0\\
630	0\\
640	0\\
650	0\\
660	0\\
670	0\\
680	0\\
690	0\\
700	0\\
710	0\\
720	0\\
730	0\\
740	0\\
750	0\\
760	0\\
770	0\\
780	0\\
790	0\\
800	0\\
};
\addlegendentry{Probe time}

\end{axis}
\end{tikzpicture}%
	\vspace{-0.5cm}
	\caption{Histograms of the Probe times and the AES encryption times of the \Sbox\ implementation measured in a machine shipped with an Intel Core i5-7600K including 1000 time samples.}
        \label{fig:probe_sbox}
\end{figure}

As opposed to the T-Table implementation, the \Sbox\ implementation does not merge different operations into one. The table is used 16 times each round, once per round input byte, during the \emph{SubBytes} operation. Considering a cache line size of 64 bytes, such table uses 4 cache lines (256 bytes). If we compute the probability of not accessing one of these cache lines and assume a key size of 128 bits, and as a result 10 rounds, such probability is equal to 0, as shown in equation~\ref{eq:pr10}. As a consequence, an attacker observing the cache before and after the encryption process will not gain any information from that observation.
{\small
\begin{equation}
\Pr[\text{no access \Sbox\ in encryption}]= \left( 1-\dfrac{64}{256} \right)^{10*16}= 0
\label{eq:pr10}
\end{equation}
\begin{equation} 
\Pr[\text{no access \Sbox\ in round}]= \left( 1-\dfrac{64}{256} \right)^{16}= 0.01
\label{eq:prRound}
\end{equation}
}
Equation~\ref{eq:prRound} shows that observing each round individually would give a $0.1$ chance of not accessing one of the lines. Instead of relying on the challenging task to stop the process after each round, the countermeasure of prefetching was applied: The OpenSSL \Sbox\ implementation includes a prefetch stage before each of the rounds. Since the 256 bytes of the \Sbox\ table map to 4 different cache lines, the encryption process only has to read 4 values to ensure the whole \Sbox\ table is loaded into the cache memory. Even when the S-Box implementation performs key-dependant memory access, the data will always be in the cache. As a result, traditional side channel attacks cannot be used to extract information from this implementation. Irazoqui et al. analyzed the OpenSSL implementation in 2017 with a tool for leakage detection and came to the same conclusion, declaring it leakage free~\cite{Irazoqui2017did}. Indeed, if we compare the time it takes to perform an encryption with the time it takes to perform a single \textit{Probe}~\cite{lastlevel} in one set, it is clear that the resolution of a \PP\ attack\footnote{The Probe includes sequential accesses to the elements of the eviction set and no cache misses that would increase the Probe times} is not high enough to retrieve fine-grained information from this implementation: As we can see in figure~\ref{fig:probe_sbox}, a probe time without any cache misses takes almost as long as the entire encryption. Even if the attacker primes the cache before the encryption starts and is able to tell exactly when the encryption starts, a single cache miss will increase the time required for probing to be greater than the total encryption time. 

The \Sbox\ implementation is not the default one in OpenSSL when an AES encryption operation is triggered from the command line. However, when using the C API of the library, a call to the function \texttt{AES\_encrypt()} will use the \Sbox . A developer wishing to use the default AES-NI instructions (as in the command line) has to explicitly indicate it. While analyzing the shared library included in Ubuntu 16.04 or CentOS 7.6 (OpenSSL 1.0.2g), we observed an additional protection. The OpenSSL implementation of AES has four different S-Boxes. If there are, for example, two processes using the library at the same time, each of them will use a different table.

We discovered this implementation is still vulnerable, since there are tiny time windows during the encryption process from which the attacker can gain valuable information. As we will show in section \ref{sec:results}, our proposal allows an attacker to observe information referring to just the last round, even in the absence of shared memory. We show how to bypass the prefetch, and perform a cross-core cache attack that recovers the secret key of this theoretically protected implementation. 

\subsection{WolfSSL RSA exponentiation}
\label{sec:rsa}

RSA is the most widely used public key cryptographic algorithm. It considers a public key $(n,e)$ where $n$ is the product of two prime numbers $p$ and $q$ that remain secret, and a private key $(p,q,d)$ where $d \equiv e^{-1}\pmod{(p-1)(q-1)}$. In order to understand the attacks only the encryption and decryption operations are relevant. For a message $m$,  the ciphertext $c$ is obtained as $c=m^e \pmod{n}$ and it is recovered with an analogous operation $m=c^d \pmod{n}$. In particular the decryption, which is the exponentiation operation using the secret key, is the target of the attack.

There are multiple ways of implementing this exponentiation~\cite{KOC199517,Gordon1998}. We will focus particularly on the square-and-multiply exponentiation, since the wolfSSL implementation is based on this. The square-and-multiply approach scans the bits of the secret exponent, performing a square operation independently of the value of the scanned bit, and a multiplication if such bit is equal to 1. Thus, an attacker monitoring the square and multiply operations can retrieve the sequence of bits of the exponent. 

The countermeasures wolfSSL has deployed to protect this implementation are to always perform the square and the multiply operations and to load the two possible values of the bit to keep them in the cache so they prevent an attacker from distinguishing which one (0 or 1) was used. These are clearly put in place to prevent cache attacks, which can be seen in both the source code comments and the release notes~\cite{Barthelmeh2016wolfSSL,OuskaSSL2016tfm}. The resulting procedure is summarized in algorithm \ref{rsawolf}. 

\renewcommand{\algorithmicrequire}{\textbf{Input:}}
\renewcommand{\algorithmicensure}{\textbf{Output:}}
\algnewcommand{\algorithmicendif}{\textbf{end if}}
\algblockdefx[IF]{If}{EndIf}[1]{\algorithmicif\ #1\ \algorithmicthen}{\algorithmicendif}
\algnewcommand{\algorithmicendfor}{\textbf{end for}}
\algblockdefx[FOR]{ForAll}{EndFor}[1]{\algorithmicfor\ #1\ \algorithmicdo}{\algorithmicendfor}
\begin{algorithm}
	\caption{wolfSSL exponentiation implementation}
	\label{rsawolf}
	\begin{algorithmic}[0]
		\Require base $b$, modulo $m$, exponent $e=(e_{n-1}...e_0)_2$
		\Ensure $b^e \pmod{m}$
		\State init$(R);$
		\For{ $i$ from $n-1$ downto $0$ }
		\State mul$(R[0],R[1],R[e_{i}]));$ \label{lin:mul}
		\State red$(R[e_{i}]);$ \label{lin:red}
		\State sqr$(R[2],R[2]);$ {\footnotesize \Comment{temp variable that avoids the leakage of $R[e_{i}]$}}
		
		\State red$(R[2]);$
		\EndFor
		\State \textbf{return} $R$;
	\end{algorithmic}
\end{algorithm}

In order to analyze the wolfSSL implementation of the exponentiation, we compiled the latest version at the time of writing this paper (version 4.4.0) with the \emph{--enable-debug --enable-keygen} flags in order to be able to keep the symbols after the installation and to generate RSA keys. Later, we ran the tests included in the library itself to analyze their RSA implementation and found that the exponentiation is still vulnerable despite the steps taken to remove side channel vulnerabilities. 

As in the case of AES, this approach still leaves a window long enough for observations of the bit values. Indeed there are two possible windows to retrieve the secret information. Firstly, at the end of the multiplication operation in line~\ref{lin:mul} only the result referring to the actually used bit is stored. In that copy process, one of the two possible values is loaded while the other one remains untouched. This leaks the key bit. The second window is even bigger, because the reduce operation in line~\ref{lin:red} only uses the information of the actual key bit value, not taking the precaution of loading both values. This means that this function could be even vulnerable to a traditional cache attack, although the synchronization between the attacker and the victim process would be a challenge.

\section{Attack scenario and overview}

One of the goals of this work is to demonstrate an attack mechanism that gives an attacker more insights about what the victim is doing and further control of the cache evictions so she can obtain the desired information. We first describe the considered scenario, in which the attacker does not interact with the victim directly. Next, we describe the steps of the \attack\ approach.

\begin{figure}
	\centering
	\includegraphics[width=0.47\textwidth]{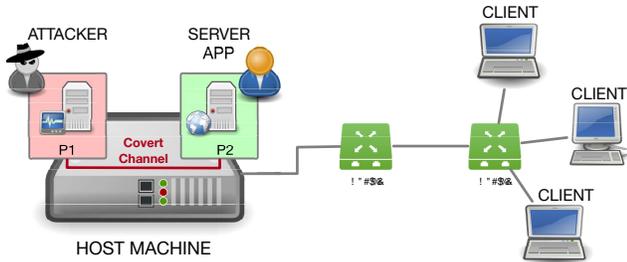}
	\vspace{-6mm}
	\caption{Diagram of the considered scenario for the attack against the \Sbox\ implementation of AES. The attacker and the victim share the hardware}
	\label{sbox:scenario}
\end{figure}

In the considered scenario, which is depicted in Figure~\ref{sbox:scenario}, we can find the following agents:

\begin{description}
	\item[Server] Encrypts/decrypts a block of data whenever it gets a request from any client. While this is a simplified version of a real server process, it is enough to create a realistic scenario for the attacks. 
	\item[Client] Sends requests at random times, each around 500$\mu$s plus a random time. This process tries to emulate the normal behavior of a real network.
	\item[Attacker] Monitors the cache to detect the exact times when the target process (an encryption process) is running so it can launch a precise attack.
\end{description}

Our main assumption is that both the attacker and the victim are using the same machine. Since we assume no synchronization between the victim and the attacker, the attacker does not know when the victim is running an encryption. Thus, the first challenge for the attacker is to detect when the victim is performing an encryption, which she does by spying on the cache. 

The second challenge for the attacker is to design a technique that allows her to evict the target data from the memory at the desired instant. Note that this eviction has to be accurate since the attacker has a very limited time window to observe the victim behavior. To retrieve necessary information up front, the attacker needs to perform an offline \emph{stakeout} phase. Then, \attack\ is conducted in three steps:

\begin{enumerate}
    \item[Aim] Detect the execution of the victim process.
    \item[Wait] Wait for a pre-determined time 
    \item[Shoot] Cause an eviction of the desired data.
\end{enumerate}

We will explain the stakeout and the aim and shoot steps in the following sections. The wait step is straight forward after completing the stakeout. Note that although we use AES for exemplification, the presented approaches are general, and can be applied to virtually any other target. The pseudocode of the attack is displayed in listing \ref{aessbox}, and all line numbers in the following refer to this code. 

\subsection{Stakeout: Preparing for attack}
The goals of the stakeout-phase are to find an appropriate cache region to monitor and determining the waiting time for the second attack step. This can be performed offline and is completely independent of any victim executions. 

For AES, we would like to monitor the \Sbox. As stated in section~\ref{subsec:aes_background}, we have to determine which of the four \Sbox es is used by the server. In our case it is the third table, and it seems to stay the same through executions. To retrieve this information, an attacker can just perform an unmodified version of the \FR\ or even \PP\ attacks and profile the process. 

The waiting time is the time between the start of the encryption and the relevant observation window. The constant \textbf{WAIT\_TIME} represents that value in listing \ref{aessbox}. One possible way to obtain it is to profile the victim application on a machine similar to the target machine. The attacker measures the time it takes the process to execute the code between the point that serves for detection and the point in which the attacker can gain information from the evicted data, for example, the last round of the AES encryption process in our scenario. Note that the time it takes to execute the \emph{clflush} has to be considered but it is not significant if no fence instructions are used to measure its execution time. As a result, we can evict data that is accessed as soon as 60 cycles after the execution of the line used for detection. This time has to be considered when selecting the gadget function.

A second approach assumes that an attacker does not know the waiting time in advance, but she knows some characteristics of the process as the probability of observing a cache hit or miss in the target. For example around 1\% of cache misses will be ideally observed if we hit exactly the last round of the AES encryption or around 50\% of the bits are expected to be 1 in an RSA secret key. In this case, the value for \textbf{WAIT\_TIME} can be retrieved automatically by analyzing the number of hits and misses observed in the recovery step (lines \ref{lin:rec} - \ref{lin:end}) and modifying its value accordingly. Even further, the attacker can define an initial value for \textbf{WAIT\_TIME} and update or adapt it dynamically based on the comparison between the actual and the expected observations. 

\subsection{Aim: Detection of the victim's execution}

In order to spy on a reduced time window during the execution of the target algorithms we must be able to identify such window during the runtime. For this reason, we study different detection mechanisms to cover a variety of attack scenarios. In the most favorable scenario, the host machine has Intel TSX, which we aim to use for detection. In the second scenario, the attacker relies on the existence of shared memory and on the knowledge of some characteristics of the target algorithm. Indeed, as we will show in later sections, if the victim prefetches the target data, we can use that prefetch for detection. Finally, and although we do not explore further in that direction we briefly describe other approaches that could be used for detection.

To evaluate which of the different approaches works better, we have modified the server, so it gives us information about the precise time instants at which the target process (an AES encryption) begins and ends. We then use this information to evaluate how many of the encryptions have been detected and if so, how many of them are detected before the encryption has ended to ensure we still have time for making an observation. 

In particular, we monitor one of the lines of the \Sbox. And since the prefetching stage of the targeted OpenSSL AES implementation involves the \Sbox\ being loaded into the cache at the beginning of the encryption process, this reliably informs the attacker when the encryption has started: whenever we detect the presence of the \Sbox\ data in the cache, we assume the encryption has started. 

\subsubsection{Leveraging TSX for detection}

As we have already stated, the process running inside a transaction can either be completed and commit the results or suffer an abort and rollback the computations. 

The \PA\ attack \cite{disselkoen17} deliberately causes conflicts in the L3 cache with the victim process to determine whether it has or has not used certain data. We use such conflicts in the L3 cache to determine the exact instants when the process executes the target instructions or uses certain data we want to monitor. The specific targets here are the data in the \Sbox\ in the attack against AES and one instruction in the vulnerable exponentiation function in the case of the attack against RSA. 

First of all we build eviction sets mapping to the same sets as our targets. We enabled hugepages to construct eviction sets as Liu et~al. did in \cite{lastlevel}, although it is possible to use the reverse-engineered mapping function \cite{Maurice_2015} or a different approach that does not require the use of huge pages ~\cite{islam2019spoiler,vila2018theory}. We read the data in one eviction set inside the transaction and wait for the victim to execute. We observed some spontaneous aborts as also noticed in previous works \cite{disselkoen17,203672}, but we were able to properly detect 97\% out of 10000 encryption processes executed during this experiment. We believe that during the ``undetected'' executions we were loading the data into the transactional region concurrently or just after the execution of the victim process, or an unrelated process accidentally evicted our data.  

Moreover, we have used the abort handler to read the system timestamp and compared this value with the timestamp we have collected in the server just before executing the encryption process. The difference between them is almost constant and the mean the value of this difference is 380 cycles in our system (Intel Core i5-7600K). Based on this measurements, we state that this technique accurately informs about the execution of the victim and that the attack gets the abort as soon as the conflict happens. Note that the measured times include the execution of some instruction before the prefetch of the \Sbox\, and the time it takes the CPU to retrieve the \Sbox\ data from the main memory. That is, the transaction aborts once the data has been effectively loaded into the cache and the attacker's data is removed. 
While a heavy system load leads to additional aborts that are not related to the encryption, it does not influence the detection time. 

\subsubsection{Shared Memory and detection}

The \FR\ ~\cite{YaromEtAl2014} technique has revealed itself as one of the most reliable sources of information in cache attacks, especially when compared to the \PP\ technique \cite{irazoqui14wait,sca,lastlevel,app9050944}. The main reason is that, in the former case, observations are made on a shared memory block, whereas in the latter case, any other process running in the machine could force an eviction and the attacker would not be able to distinguish the origin of the eviction. The \FR\ technique requires the victim and the attacker to share the library. This is often not a problem since, for example, we have observed that both Ubuntu 16.04 and CentOS 7.6 are shipped with a compiled version of \textit{libcrypto} (OpenSSL 1.0.2g) and newer versions of Ubuntu  such as Ubuntu 20.04 also come with the latest version of the shared library (OpenSSL 1.1.1f). Similarly, wolfSSL, when compiled and installed, generates a shared library. 

Back to the \Sbox\ example, the attacker flushes the data of the \Sbox\ line monitored in the previous scenario and then waits for an arbitrary time and reloads that data. We have observed that, if we do not include this waiting time, we are not able to observe any cache access. We have similarly considered two different ways of ensuring that the process waits for a fixed time: increasing the value of a variable up to a limit or actively polling the system counter through the \texttt{rdtsc} instruction and constantly checking if we have waited for the desired time. Each approach yields to different results, but they are similar in the sense that they lead to similar detection rates.

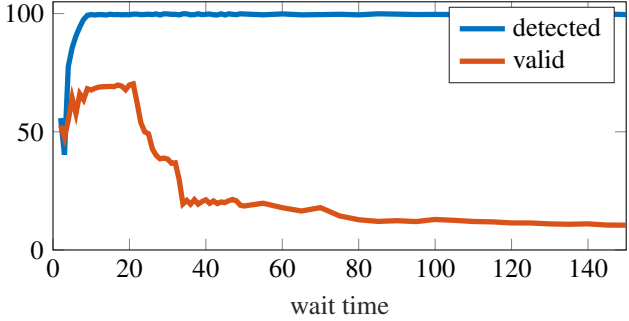
\begin{figure}
	\centering
%
%
\definecolor{mycolor1}{rgb}{0.00000,0.44700,0.74100}%
\definecolor{mycolor2}{rgb}{0.85000,0.32500,0.09800}%
\begin{tikzpicture}

\begin{axis}[%
width=3in,
height=1.3in,
at={(1.372in,0.566in)},
scale only axis,
xmin=0,
xmax=150,
xlabel style={font=\color{white!15!black}},
xlabel={wait time},
ymin=0,
ymax=105,
axis background/.style={fill=white},
legend style={legend cell align=left, align=left, draw=white!15!black}
]
\addplot [color=mycolor1, line width=2.0pt]
  table[row sep=crcr]{%
2	55.797655\\
3	40.268348\\
4	77.65\\
5	85.469355\\
6	90.505928\\
7	94.340513\\
8	97.499147\\
9	99.274305\\
10	99.58969\\
11	99.419391\\
12	99.619454\\
13	99.526255\\
14	99.437186\\
15	99.726749\\
16	99.559945\\
17	99.609531\\
18	99.505458\\
19	99.571841\\
20	99.51338\\
21	99.767542\\
22	99.776501\\
23	99.559945\\
24	99.631364\\
25	99.749628\\
26	99.597626\\
27	99.863187\\
28	99.42631\\
29	99.899102\\
30	99.865182\\
31	99.671085\\
32	99.718793\\
33	99.458943\\
34	99.914074\\
35	99.892117\\
36	99.492588\\
37	99.562919\\
38	99.908085\\
39	99.549043\\
40	99.694934\\
41	99.826302\\
42	99.519322\\
43	99.42631\\
44	99.914074\\
45	99.431253\\
46	99.918067\\
47	99.69394\\
48	99.561928\\
49	99.926055\\
50	99.834275\\
55	99.495558\\
60	99.868174\\
65	99.476752\\
70	99.582748\\
75	99.71581\\
80	99.484669\\
85	99.912077\\
90	99.780483\\
95	99.608538\\
100	99.656186\\
105	99.59961\\
110	99.975006\\
115	99.717799\\
120	99.96801\\
125	99.64427\\
130	99.986002\\
135	99.547061\\
140	99.932046\\
145	99.839259\\
150	99.578782\\
};
\addlegendentry{detected}

\addplot [color=mycolor2, line width=2.0pt]
  table[row sep=crcr]{%
2	52.58585\\
3	47.633924\\
4	54.959998\\
5	64.311545\\
6	57.732341\\
7	66.020001\\
8	63.562344\\
9	68.215394\\
10	67.673643\\
11	68.40251\\
12	68.915335\\
13	69.078335\\
14	69.115042\\
15	69.20436\\
16	69.122034\\
17	69.799215\\
18	69.34365\\
19	67.799698\\
20	69.823462\\
21	70.277088\\
22	62.288568\\
23	53.787704\\
24	49.98148\\
25	49.156359\\
26	42.783784\\
27	40.018421\\
28	38.573259\\
29	38.884385\\
30	38.467082\\
31	36.70183\\
32	36.885459\\
33	29.915317\\
34	19.494935\\
35	20.978398\\
36	19.373793\\
37	21.286395\\
38	19.386029\\
39	20.488488\\
40	21.216246\\
41	19.731112\\
42	20.752753\\
43	19.624568\\
44	20.314127\\
45	20.047236\\
46	20.844636\\
47	21.353061\\
48	20.781028\\
49	18.955576\\
50	18.610983\\
55	19.773496\\
60	17.885072\\
65	16.488248\\
70	17.915469\\
75	14.417617\\
80	12.788136\\
85	12.090388\\
90	12.389469\\
95	12.019067\\
100	12.912402\\
105	12.536954\\
110	12.050594\\
115	11.896477\\
120	11.4371\\
125	11.41832\\
130	11.034282\\
135	10.879462\\
140	11.053497\\
145	10.532925\\
150	10.501992\\
};
\addlegendentry{valid}

\end{axis}
\end{tikzpicture}%
	\vspace{-0.5cm}
	\caption{Percentage of correctly detected encryptions and valid encryptions for an attack as a function of the waiting time measured as the limit of a count}
        \label{wait_flush}
\end{figure}

Figure~\ref{wait_flush} shows the results in terms of correctly detected encryptions and valid encryptions (the encryption was detected before the victim had actually finished it) as a function of the waiting time. The waiting time refers to the limit up to which a variable is increased: from 0 to the limit. As we did in the previous case, we collect information referring to 10000 encryptions per each considered limit. As it can be derived from the figure, limits around 20 obtain the best results in terms of valid encryptions. As a consequence, an attacker should select that limit to maximize the chances of success when trying to evict data later during execution of the target window. In any case, these results show that it is also possible to use the \FR\ technique for detection in case the processor does not include TSX, although the results will not be as accurate as the ones obtained using TSX. 

\subsubsection{Other scenarios}

Another option to detect the execution of the victim process are \PP\ based approaches. If we look back to figure \ref{fig:probe_sbox} and compare the Probe times with the execution of the whole encryption process it seems clear that performing a complete Probe of the data in the eviction set is not a real option to obtain valid information for the attack from a single round. However, considering the replacement policy of the L3 cache and that the attacker knows the insertion order of the elements in the eviction set, she can sequentially access such elements from the one that was first inserted to the one that was last inserted. This way, she can be reasonably sure that she is accessing the element that the target will replace in case of conflict.

The main drawback of this approach is that even if the detection is successful, the attacker does not have any control of the state of the cache at the time the victim executes. As a result, she cannot accurately predict the number of accesses that will be required to evict the recently inserted data. Since the capability of the attacker to achieve this eviction during the execution of the victim process is determined by this state of the cache and the number of required accesses, the attacker capabilities are rather limited in this scenario.

\subsection{Shoot: Accurate eviction of the data}

Due to the prefetching or to the always execute and always load into the cache strategies, the ability of the attacker to retrieve the secret information is determined by his ability to evict the data from the cache during the vulnerable windows. Namely, when attacking the \Sbox\ implementation of AES our goal is to achieve evictions during the last round and after the last prefetch, when attacking the RSA exponentiation, our target are certain windows during the execution of some operations in the multiply or the reduce functions.

\begin{figure}
	\centering
	\begin{tikzpicture}[>=stealth', 
	decision/.style={draw, font=\scriptsize, text width=1.8cm, align=center, inner sep=1pt, minimum width=2.1cm, diamond, aspect=1.6},
	goal/.style={draw, font=\scriptsize, rectangle, rounded corners=5pt, align=left, text width=2.3cm, node distance = 2.9cm}, 
	arr/.style={->,thick, fill}
	]
	\node[decision] (tsx) {Is TSX available?};
	\node[decision, inner sep = -1pt] (sm1) [below of = tsx, node distance = 1.8cm] {Is there shared memory?};
	\node[decision, inner sep = -1pt] (sm2) [right of = tsx, node distance = 2.9cm] {Is there shared memory?};
	\node[goal] (goal1) [right of = sm2] {AIM: \FR\  \\ WAIT(WAIT\_TIME) \\ SHOOT: Flush};
	\node[goal] (goal2) [right of = sm1, fill=blue!10] {AIM: TSX (abort)  \\ WAIT(WAIT\_TIME) \\ SHOOT: single access};
	\node[goal] (goal3) [below of = sm1, node distance = 1.6cm, fill=blue!10] {AIM: TSX (abort)  \\ WAIT(WAIT\_TIME) \\ SHOOT: Flush};
	\draw[arr] (tsx) -- (sm1) node[left, midway] {\scriptsize{Yes}};
	\draw[arr] (tsx) -- (sm2) node[above, midway] {\scriptsize{No}};
	\draw[arr] (sm1) -- (goal3) node[left, midway] {\scriptsize{Yes}};
	\draw[arr] (sm1) -- (goal2) node[above, midway] {\scriptsize{No}};
	\draw[arr] (sm2) -- (goal1) node[above, midway, xshift=-2pt] {\scriptsize{Yes}};
	\end{tikzpicture}
	\caption{Flow diagram of the possible scenarios that enable the proposed attack}
	\label{eviction_scenario}
\end{figure}
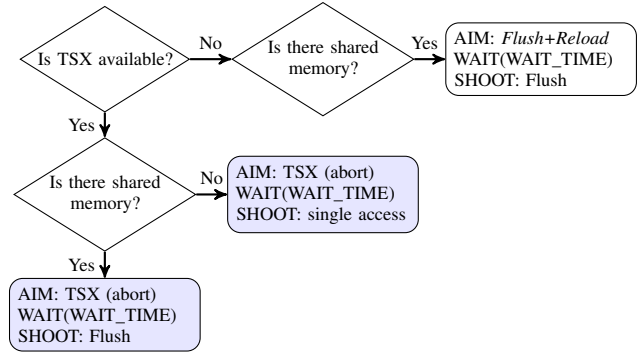

As we did when explaining the detection scenarios, we also consider different possibilities in this case. The considered scenarios are depicted in figure \ref{eviction_scenario}. Note that since we obtained the best detection results for the machines with TSX, this is the feature we first look for, although it will not be used during the eviction process. In figure \ref{eviction_scenario} we have highlighted the scenarios we are going to describe more deeply in this section for the shake of simplicity and brevity. In all cases the idea is the same, once we know the victim is running we wait for the exact time determined in stakeout, which will allow us to force the eviction at the desired instant.

In the following, we refer to the technique that allows us to evict data from the cache once the execution of the victim has been detected using the TSX capabilities as method 1 when there is shared memory between the victim and the attacker, and as method 2 when there is not. Although not explicitly mentioned in the figure, the attacker has to recover the information leaked thanks to the accurate eviction of the data. For method 1, reloading the shared data is enough, in the second case, another access is required.

In order to help us with the explanation of both methods, we include the pseudocode of the \attack\ attack process (Aim, wait, shoot, then recover information) particularized to the \Sbox\ attack that we have been using as example in this section. Note that the procedure procedure summarized in the algorithm~\ref{aessbox} is general, the attacker just has to change the \Sbox\ by her target, for example the RSA multiply function in our other example, and adapt the waiting times and the eviction function correspondingly. 

As we will show, method 1 differs from method 2 in the way the cache set is filled during the transaction, in the value of the \textbf{WAIT\_TIME} even if the target is the same, in the way this target is evicted from the cache and finally in the way the information about the actual access to the data is inferred. Next, we explain the particularities of each approach.

\renewcommand{\algorithmicrequire}{\textbf{Input:}}
\renewcommand{\algorithmicensure}{\textbf{Output:}}
\algblockdefx[IF]{If}{EndIf}[1]{\algorithmicif\ #1\ \algorithmicthen}{\algorithmicendif}
\algblockdefx[FOR]{ForAll}{EndFor}[1]{\algorithmicfor\ #1\ \algorithmicdo}{\algorithmicendfor}
\begin{algorithm}
	\caption{Attack against the OpenSSL AES \Sbox\ implementation for the TSX-based detection scenario}
	\label{aessbox}
	\begin{algorithmic}[1]
		\Require \textbf{Address(S-Box(0)),Eviction\_set}  \Comment{Address of the \Sbox}
		\Ensure \textit{\textbf{$\overline{X}_{0},\overline{S}^{N_r+1}$}} \Comment{Information about the access and ciphertext}
		\For{$t=0$ \textbf{to} \emph{number\_of\_encryptions}}
		\State \textit{Start\_Transaction();}
		\If {Successfully Started}
		\State fill\_cache\_set(); \label{lin:fill}
		\Else  \Comment{Aim: Abort handler detects the access}
		\State \textbf{time\_interrupt}=\textit{timestamp()}+\textbf{WAIT\_TIME}; \label{lin:wait}
		\State \textbf{While}(\textit{timestamp()} $\leq$ \textbf{time\_interrupt}) \{\};
		\State \textbf{Evict from cache} \textbf{(S-Box(0))}; \label{lin:ev}
		\State \textit{\textbf{Wait} until encryption ends}
		\State \textbf{Infer victim access to}(\textbf{(S-Box(0))} \label{lin:rec} 
		\If {hasAccessed(\textbf{(S-Box(0))}}
		\State $\overline{X}_{0}\left[t\right]=1;$ \Comment{Data used}
		\Else
		\State $\overline{X}_{0}\left[t\right]=0;$
		\EndIf \label{lin:end}
		\EndIf
		\EndFor	
		\State \textbf{return} $\overline{X}_{0},\overline{S}^{N_r+1}$;
	\end{algorithmic}
\end{algorithm}

\subsubsection{Method 1}

As we have already stated, this method relies on the existence of shared memory. That allows the attacker to flush from the cache the desired data when the desired instant comes. Once the transaction is started, the attacker has to fill the cache set (line \ref{lin:fill} in the algorithm \ref{aessbox}). It is not required for the attacker to place the data in any specific order but she must fill the whole set to get the abort signal that will inform her that the victim is accessing the target. One can see this detection mechanism as a ``break point'' the attacker places into the code without notifying it to the victim.

Once the abort is triggered by the victim's access, we use the handler to carry the attack. This process includes mainly the definition of the time the attacker has to wait (line \ref{lin:wait}), the procedure to evict the data (line \ref{lin:ev}) and finally a way to retrieve and store the desired information (lines \ref{lin:rec} - \ref{lin:end}). Since there is shared memory data is evicted using the \emph{clflush} instruction and retrieved measuring the time it takes to read such data (reload). This is a traditional \FR\ attack carried by an abort handler, which allows for very precise timing of the flush instruction. 

\subsubsection{Method 2}

If there is no shared memory or the data is going to be retrieved from a non shared variable (e.g. in the attack against RSA), it is still possible to achieve the desired eviction accurately by just accessing one memory location. In order to be able to do this, we have to manipulate the cache state prior to the detection of the encryption process. Indeed, the required state of the cache, depends on the replacement policy of the last level cache and it can be changed by accessing data located in the cache. Note that accesses to data located in lower caches do not change the state of the last level cache. Interestingly, the state of the cache remains unchanged when the transaction aborts. In other words, the abort does not revert the microarchitectural state of the cache. 

One of the simplest ways to check this assumption, is flushing some data out of the cache before executing the transactional code. Then, in the transactional region, reload that data back into the cache. Next, we wait in an endless loop doing nothing within the transaction, until an abort happens. The abort handler checks if the data is in the cache or not. We have repeated this test 10000 times, and the conclusion is that the data that has been loaded into the cache during a transactional operation, remains in the cache after such operation has finished, in the sense that even if the process never gets to see the result of the code executed inside the transaction, the microarchitectural state of the processor retains some information. 

This is similar to transient execution attacks ~\cite{Kocher2018spectre,217478}, which exploit the fact that microarchitectural changes of transient executions that have never been committed are still observable from the architectural state. 

Although the manipulation of the data in the cache is performed during the preparation of the code for the detection, in the fill\_cache\_set() function of the algorithm~\ref{aessbox} (line \ref{lin:fill}), it is directly related with the eviction, and this is the reason why we are giving more details in this section. This initial setup is also one of the differences between method 1 and method 2. Also, since this approach evicts the target by accessing another block instead of using the \emph{clflush} instruction (line \ref{lin:ev}) this has an effect on the \textbf{WAIT\_TIME} value. Namely, the target data is not evicted from the cache until the replacement block has been retrieved from the main memory, and this introduces a delay of around 200 cycles since detection. This means that around 250 cycles must elapse after detection, otherwise it is not possible to evict the data. The profiling of the victim for the selection of the \textbf{WAIT\_TIME} or the automation of the process to obtain it is still necessary. The last difference to method 1 is how data is retrieved, in this case another access is required (line \ref{lin:rec} )

In the following, we give more information about why it is possible to use TSX to support our attack, how this approach works and some considerations related with the L1 cache replacement policy.

\paragraph{Replacement policy and accurate eviction of data}\mbox{} \\

The replacement policy described in \cite{Briongos2019} agrees with the ones described in \cite{vila2019cachequery,Abel19} in the selection of the eviction candidate most of the times. Indeed, our attack can be explained considering any of them as the correct one. Mainly, once the data is inserted into the cache set, all of them distinguish two different ages above the age 0, and they all consider that the data is evicted from the cache with age 3. They differ in the way the ages are updated. Since our attack was originally designed using the approach described in \cite{Briongos2019}, we will use that policy for the explanation.

Besides the replacement policy, there are two things we need to bear in mind to understand this approach. The first one is that only access to the last level cache update the values of the ages of the elements in the LLC. The second one is that the LLC in our machine is inclusive. This means that if the victim loads the \Sbox\ into the cache after a cache miss, it will be loaded in all the cache levels, and in case this data is used multiple times during the encryption, it will be retrieved from the low level caches leaving the age of the cache line holding the \Sbox\ unchanged. As a result, if it is the oldest element in the set when inserted, it will be evicted in case of conflict.

\begin{figure}
	\centering
	\includegraphics[width=\columnwidth]{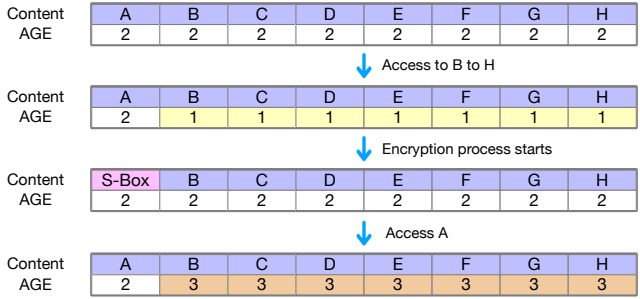}
	\vspace{-0.5cm}
	\caption{General diagram of the process required to evict the data from the cache with just one access.}
	\label{ev_process}
\end{figure}

The general idea of the attack is depicted in Figure~\ref{ev_process} particularized for the \Sbox\ attack. As a difference with the previous approach the attacker needs to set the ages of the elements that have just been inserted into the set (stage 1) to the ones shown in the second stage of the figure. This means that the \texttt{fill\_cache\_set()} function in the algorithm ~\ref{aessbox} involves two steps. Note that all the elements in the eviction set are accessed, so the transaction still aborts when the victim accesses the monitored data. 

Once the encryption begins, the \Sbox\ data will be placed into the cache. In that moment, the oldest element (A) that we have not accessed to ensure it is older than the rest of elements, will be evicted. At the moment the \Sbox\ data is inserted, and considering all the updates in the ages of the blocks, all the blocks (the ones originally placed in the cache by the attacker and the \Sbox\ block) will have the same age. Since we perform the accesses in such a way that the oldest element before inserting the \Sbox\ data (A) is located in the first position in the set, the \Sbox\ automatically becomes the eviction candidate just after being inserted. 

Then the procedure is similar as it was in the previous case. We wait for the desired time and then by accessing A (which is out of the cache because it was replaced with the \Sbox\ data) we ensure that the data of the \Sbox\ is removed from the cache. We wait until the encryption ends, and finally retrieve the information about the actual access to the \Sbox\ by accessing B. Since B is the first element with age 3, in case of conflict it will be replaced with the \Sbox\ data. Note that after the recovery step we need to change the ages again in a similar way they did in the Reload+Refresh attack \cite{Briongos2019}.

\paragraph{Influence of the replacement policy of the L1 caches}\mbox{} \\

When trying to force all the elements in the set to get the ages depicted in the stage 2 of Figure~\ref{ev_process}, we accessed the corresponding blocks B to H as a linked list (to avoid pipeline effects as much as possible). We thought that this way all the data would be retrieved from the LLC and as a result, the ages of all the elements would be updated. To test our hypothesis that the data was retrieved from the LLC, we measured the times it took us to read each of the blocks. Surprisingly, we found out that some of them were retrieved from the L1 cache. The replacement policy of L1 and L2 caches was responsible for this effect, so we performed some experiments to understand how it works.

In fact, we only observed this behaviour in our processor LLC whose LLC is 12-way associative, whereas the L1 and the L2 caches are 8-way and 4-way associative respectively. However, we performed the same test in a different processor (Intel i7-6700K) with a 16-way associative cache, and all the data was retrieved from the LLC. In the first case, just after the 12 elements of the set have been placed in the cache, 4 of them are only present in the LLC, and the L1 cache will have suffered 4 misses but still keep 4 of these blocks. The replacement policy defines which elements are replaced and which are replaced. On the contrary, if the number of ways of the LLC is 16, the L1 cache will have suffered 8 misses. This means it is likely that the 8 elements that were first accessed only reside in the LLC. We observed that in the first case the block we call B is in the L1 cache whereas it is not in the second case. For these reasons we conducted some experiments to determine the replacement policy in the low level caches. 

Low level caches are assumed to implement a Pseudo-LRU replacement policy, simpler than the policy implemented in the LLC. We conducted some experiments aimed to retrieved that policy. Based on our observations and some intuition we already had, we were similarly able to explain our results assuming a replacement policy and checking times. It turns out that Intel implements a tree based Pseudo-LRU replacement policy in the low level caches. Independently and concurrently with the design of this attack, some other researchers arrived to the same conclusion \cite{vila2019cachequery,Abel19}.

\begin{figure}
	\centering
	\includegraphics[width=0.95\columnwidth]{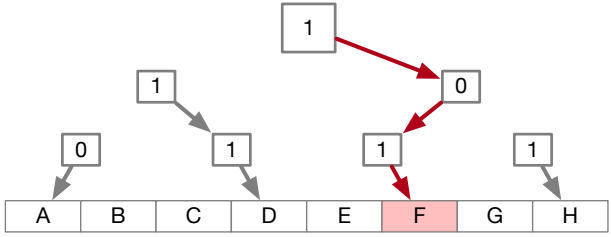}
	\vspace{-0.2cm}
	\caption{Tree structure that controls the Pseudo-LRU replacement policy of L1 and L2 caches. The eviction candidate is highlighted}
	\label{tree_lru}
	\vspace{-3mm}
\end{figure}

To sum up, the tree-based replacement policy is represented in Figure~\ref{tree_lru}. Starting from the root node it selects each of the branches depending on the intermediate values of the nodes. In the example the eviction candidate is marked in red. Since the root node contains a 1, it selects the right branch. The value of the pointed node is a 0, so it selects the left branch. Finally, the last node has a 1, so it points to the element at the right, F in the example. Note that the blocks of memory in the cache set (A to H) are ordered. According to our experiments when the cache set is completely empty, the elements are inserted linearly in the first free block they find regardless of the actual values of the nodes. Once the set gets completely filled with data, the apparent value of all the nodes seems to be 0. If an element in the cache is either accessed or replaced, the values of the nodes that pointed to it are switched.

This replacement policy explains why B is in the L1 cache after reading the whole LLC eviction set (12 elements). For this reason, B cannot be the first element to be accessed because its age would not change. Based on this replacement policy, we have prepared a linked list of addresses to access 11 elements of the eviction set out of 12 possible in our 12-way cache, that ensures all of them are in the LLC. This way we are able to reduce their age and indeed, we have checked that the designed access pattern retrieves all the elements in that list are retrieved from the LLC.

\section{Results}
\label{sec:results}

In this section we explain how we recover the AES and RSA secret keys, demonstrating that the countermeasures to prevent cross-core last level cache attacks can be circumvented. In the cases where some adaptations of the proposed approach are possible, we mention it. All the experiments were performed in the machine described in table~\ref{tab:setup1}. The version of OpenSSL used is the one included in our CentOS system (OpenSSL 1.0.2k). Note that the replacement policy on which this attack relies is implemented in the Intel Core processors starting from the 6th generation.

\begin{table}[ht]
    \centering
    {\small{
	
	\vspace{-0.5cm}
	\caption{Experimental platform details. \label{tab:setup1}}
	\begin{tabular}{ll}
			\hline
			\textbf{Processor} & Intel core i5-7600K \\
			\textbf{Cores} & 4\\
			\textbf{Frequency} & 3.8 GHz \\
			\textbf{LLC slices} & 8\\
			\textbf{LLC size} & 6 MB\\
			\textbf{LLC ways} &12 \\ 
			\textbf{L1 size} & 32 KB\\
			\textbf{L1 ways} & 8 \\ \hline
	\end{tabular}
	}
    }
	\vspace{-0.5cm}
\end{table}

\subsection{Attack against the AES}

This targeted \Sbox\ implementation was also attacked in~\cite{DanielCachezoom}, albeit for a much more powerful adversary with full OS control targeting SGX. That work monitored the entire L1 data cache, observing various samples per round, allowing them to distinguish the prefetching stages from the normal operations of the round. Our approach on the contrary does not need to frequently interrupt the victim, works across cores and does not have any special requirement, just user-level privileges as commonly assumed for cache attacks.

As we have already stated in the previous sections of this work, the target of our attack against the \Sbox\ is the last round of the encryption. In this round, the contents of the \Sbox\ is xored with the corresponding round key to get the ciphertext. In order to retrieve the secret key in all scenarios (method 1 and 2) we use the information referring to the accesses to the \Sbox\ retrieved during the attack phase (Algorithm~\ref{aessbox}) and assume the ciphertext to be known by the attacker. 

We use the \textit{non-access} approach described in \cite{app9050944}. In a nutshell, they used information from cache misses, that is whenever the victim did \emph{not} load the data into the cache. Note that in this particular implementation, the \Sbox\ is accessed 16 times during the last round, and even if we are able to accurately get the information referring to the last round exclusively, we would not know which of the 16 accesses was responsible for this access. On the contrary, if we determine that an element has not been accessed it means none of the operations in the last round has used it. As a consequence, we xor each byte of the ciphertext with the 64 values of the \Sbox\ held in the cache line ($k_{i}= C_{i} \oplus$ \Sbox\ [0 to 63]). None of these values could be the secret key.

\begin{figure}
	\centering
	\vspace{-0.3cm}
%
%
\definecolor{mycolor1}{rgb}{0.00000,0.44700,0.74100}%
\begin{tikzpicture}

\begin{axis}[%
width=2.7in,
height=1.3in,
scale only axis,
xmin=0,
xmax=16,
xlabel style={font=\color{white!15!black}},
xlabel={Operation at which the data is evicted from the cache},
ymin=0,
ymax=80,
ylabel style={font=\color{white!15!black}},
ylabel={Probability},
axis background/.style={fill=white}
]
\addplot [color=mycolor1, only marks, forget plot]
  table[row sep=crcr]{%
1	1.00225957576185\\
2	1.33634610101581\\
3	1.78179480135441\\
4	2.37572640180588\\
5	3.16763520240784\\
6	4.22351360321045\\
7	5.63135147094727\\
8	7.50846862792969\\
9	10.0112915039062\\
10	13.348388671875\\
11	17.7978515625\\
12	23.73046875\\
13	31.640625\\
14	42.1875\\
15	56.25\\
16	75\\
};
\end{axis}

\begin{axis}[%
width=3in,
height=1.3in,
at={(0in,0in)},
scale only axis,
xmin=0,
xmax=1,
ymin=0,
ymax=1,
axis line style={draw=none},
ticks=none,
axis x line*=bottom,
axis y line*=left
]
\end{axis}
\end{tikzpicture}%
	\vspace{-0.5cm}
	\caption{Probability of not accessing a line during the last round of the encryption depending on the number of instructions of the last round that have been actually executed.} \label{aes_last}
	\vspace{-0.3cm}
\end{figure}
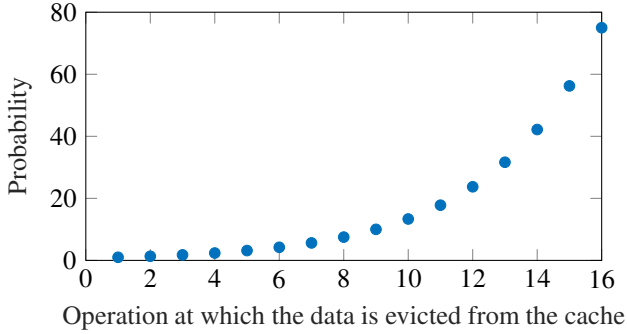

In our test system the last round takes around 50-60 cycles to execute. However the encryption time is not constant, and has a variance of around 15 cycles. Thus, even if we interrupt the victim at the exactly chosen time, which is the estimated time when the last round starts to perform its operations, we may not hit our target. Even slight variation in the interruption instantly lead to a different number of observed accesses. If we interrupt before the last prefetch, we will only observe cache hits. In contrast, if we evict the \Sbox\  data after the encryption has ended, we will only see cache misses. Indeed, assuming that the 16 operations that use the \Sbox\ in the last round of the encryption are executed sequentially from 0 to 15, the expected cache miss probability depends on the exact operation at which is evicted from the cache. This probability is depicted in Figure~\ref{aes_last}. 

To deal with the variation 
in the execution time of the algorithm and to maximize our success chances, when estimating the \textbf{WAIT\_TIME} in the Algorithm \ref{aessbox} instead of using the probability of 1\% as the expected one, we allow up to 7\% of cache misses. This way we try to ensure that the observed cache misses actually happen in the round. This value was determined empirically in our machine, by selecting a different probability values and carrying multiple experiments with the value of \textbf{WAIT\_TIME} adapted dynamically according to that probability and window sizes of $10 000$ observations.

\begin{figure}[th]
	\centering
	\hspace{-1cm}
	\input{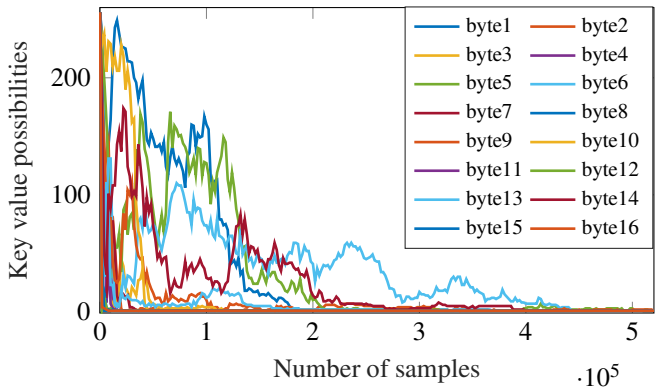}
	\vspace{-0.2cm}
	\caption{Key candidates for each of the bytes of the key of the \Sbox\ AES implementation of OpenSSL retrieved using the TSX-based detection and method 2.} \label{flush_count}
\end{figure}

Both approaches, the one that assumes shared memory (method 1) and the one that does not (method 2), are able to successfully retrieve the secret AES key. Even with the false positives (cache misses that do not refer to the key byte) introduced by all the variances, we get enough information from all the rounds. Indeed, as our results show, it is more likely to evict the data in the middle of the execution of the last round than just at the beginning. As a result some bytes are recovered faster than others as it can be observed in Figure~\ref{flush_count}. The number of samples required to retrieve the secret key slightly varies between executions and depends on the threshold selected, even if we use the adaptive approach. In both cases, the minimum number of samples required to retrieve the whole key is about \num{500000}.  

As can be observed in Figure~\ref{flush_count}, half of the key has been completely leaked after \num{100000} samples (the initial four bytes are obtained with \num{10000}  samples, although it cannot be clearly observed in the figure). Around 12 of the 16 bytes are already known with \num{200000} samples. Retrieving the last 4 bytes of the key is the hardest part, it takes \num{300000} more samples, which gives an idea about the difficulty of evicting the data in between the execution of the prefetch and the subsequent access in the last round. 

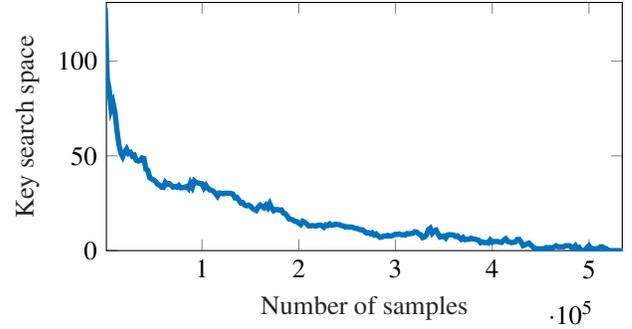
\begin{figure}
	\centering
%
%
\definecolor{mycolor1}{rgb}{0.00000,0.44700,0.74100}%
\begin{tikzpicture}

\begin{axis}[%
width=2.7in,
height=1.3in,
at={(1.255in,0.591in)},
scale only axis,
xmin=1117.31843575422,
xmax=534450.651769088,
xlabel style={font=\color{white!15!black}},
xlabel={Number of samples},
ymin=0,
ymax=131,
ylabel style={font=\color{white!15!black}},
ylabel={Key search space},
axis background/.style={fill=white}
]
\addplot [color=mycolor1, line width=2.0pt, forget plot]
  table[row sep=crcr]{%
1	128\\
2001	90.747494582851\\
4001	85.2913889023693\\
6001	74.7770846457647\\
8001	77.942629684599\\
10001	72.999293787537\\
12001	63.7135891453299\\
14001	56.0783742150433\\
16001	51.4536298571939\\
18001	49.5390822697796\\
20001	52.1019690048875\\
22001	53.6046220620714\\
24001	51.2833632523518\\
26001	51.7833666572232\\
28001	49.8638615205893\\
30001	50.3467263993231\\
32001	47.6971934115497\\
34001	47.2750072787871\\
36001	47.5185362634392\\
38001	48.9348260085333\\
40001	48.6171767515823\\
42001	42.9708977408885\\
44001	42.3298759491083\\
46001	38.462185607944\\
48001	37.7684081262701\\
50001	37.2540337145433\\
52001	36.4779917850287\\
54001	35.0322505383087\\
56001	34.5865210252865\\
58001	33.5001999872579\\
60001	33.3598400418244\\
62001	36.0039804784803\\
64001	34.9485445257516\\
66001	35.3677276595982\\
68001	34.5292313868166\\
70001	33.5565206792898\\
72001	33.8697956887699\\
74001	33.4862622140264\\
76001	34.4072837071257\\
78001	33.1479846309605\\
80001	33.1881754162264\\
82001	33.4729652025765\\
84001	33.8392713989759\\
86001	32.7500203711298\\
88001	35.5809737230144\\
90001	33.5159989118967\\
92001	36.7256188416256\\
94001	35.8743568088738\\
96001	35.7164978124015\\
98001	35.4099413374816\\
100001	35.2849377380246\\
102001	33.0940141303099\\
104001	34.4998351842116\\
106001	32.9330526545176\\
108001	32.0044651050466\\
110001	31.8891421934836\\
112001	31.210319080314\\
114001	30.1139506858418\\
116001	28.7182085511719\\
118001	30.3860077040107\\
120001	30.2530643770356\\
122001	29.9889830986206\\
124001	30.3249180515018\\
126001	30.2705405052716\\
128001	30.2927601038204\\
130001	30.2180898853846\\
132001	29.7997470229922\\
134001	27.8916747888306\\
136001	28.1382641431852\\
138001	26.7614546526316\\
140001	25.1900874940355\\
142001	25.3437159822943\\
144001	24.0611153692458\\
146001	23.6494368820568\\
148001	24.0335547421099\\
150001	23.5220532235528\\
152001	22.3500284069052\\
154001	21.7374800745121\\
156001	21.1920459379776\\
158001	22.9012440324674\\
160001	24.0347640422988\\
162001	23.2935614552461\\
164001	22.5994625106918\\
166001	24.0044083964504\\
168001	22.5846102372495\\
170001	24.756845399379\\
172001	22.5509496501098\\
174001	21.0549540121208\\
176001	21.6149398473107\\
178001	21.708598954525\\
180001	21.3215758314157\\
182001	21.4704205656901\\
184001	19.894918394218\\
186001	19.5592874570862\\
188001	18.2667865406949\\
190001	16.7021726853655\\
192001	16.5216004397237\\
194001	16.1516508299734\\
196001	15.5077946401987\\
198001	15.403544678202\\
200001	14.486835021563\\
202001	13.7448338374995\\
204001	15.4948555844912\\
206001	15.0310119074377\\
208001	14.1699250014423\\
210001	13.0768155970508\\
212001	13.0402897210257\\
214001	13.2288186904959\\
216001	13.1163439612375\\
218001	12.9068905956085\\
220001	13.3663222142458\\
222001	13.5142209093581\\
224001	13.103287808412\\
226001	12.3619437737352\\
228001	13.5919897676151\\
230001	13.2252074369435\\
232001	14.0823153941982\\
234001	13.4220647661728\\
236001	13.497352893477\\
238001	14.007027266894\\
240001	13.7846348455575\\
242001	13.7052003780631\\
244001	13.1736771363034\\
246001	13.0647427647503\\
248001	12.4304525516655\\
250001	12.6211361132746\\
252001	12.4304525516655\\
254001	12.4614794472862\\
256001	12.2585660338899\\
258001	11.7813597135247\\
260001	11.7481928495895\\
262001	11.3331553503106\\
264001	10.9943534368589\\
266001	10.9943534368589\\
268001	10.1241213118292\\
270001	9.28540221886225\\
272001	9.49185309632967\\
274001	8.90689059560852\\
276001	9.4178525148859\\
278001	8.28540221886225\\
280001	8.8073549220576\\
282001	7.8073549220576\\
284001	7\\
286001	7.16992500144231\\
288001	7.70043971814109\\
290001	7.70043971814109\\
292001	8.04439411935845\\
294001	7.62935662007961\\
296001	7.75488750216347\\
298001	8.16992500144231\\
300001	8.6724253419715\\
302001	8.49185309632967\\
304001	8.81378119121704\\
306001	8.39231742277876\\
308001	8.39231742277876\\
310001	8.16992500144231\\
312001	9.02236781302845\\
314001	8.90689059560852\\
316001	8.97727992349992\\
318001	9.16992500144231\\
320001	9.97727992349992\\
322001	9.75488750216347\\
324001	8.83289001416474\\
326001	8.4093909361377\\
328001	6.84549005094437\\
330001	7.12928301694497\\
332001	8.22881869049588\\
334001	11.0768155970508\\
336001	11.813781191217\\
338001	9.22881869049588\\
340001	10.4304525516655\\
342001	10.8454900509444\\
344001	8.8073549220576\\
346001	7.16992500144231\\
348001	7.39231742277876\\
350001	7.71424551766612\\
352001	6.6724253419715\\
354001	8.24792751344359\\
356001	8.4093909361377\\
358001	8.32192809488736\\
360001	8.22881869049588\\
362001	7.49185309632967\\
364001	7.78135971352466\\
366001	6.78135971352466\\
368001	5.90689059560852\\
370001	6.90689059560852\\
372001	5.78135971352466\\
374001	5.90689059560852\\
376001	6.02236781302845\\
378001	6.56985560833095\\
380001	6.22881869049588\\
382001	6.02236781302845\\
384001	5.78135971352466\\
386001	5.49185309632967\\
388001	4.16992500144231\\
390001	4.58496250072116\\
392001	4.16992500144231\\
394001	4.32192809488736\\
396001	4\\
398001	5.58496250072116\\
400001	4.39231742277876\\
402001	4.8073549220576\\
404001	4.8073549220576\\
406001	4.8073549220576\\
408001	4.39231742277876\\
410001	4.39231742277876\\
412001	5.58496250072116\\
414001	6.2667865406949\\
416001	4.8073549220576\\
418001	4.58496250072116\\
420001	4.16992500144231\\
422001	4.39231742277876\\
424001	4.58496250072116\\
426001	5.90689059560852\\
428001	5.90689059560852\\
430001	4.32192809488736\\
432001	2.58496250072116\\
434001	3\\
436001	3.58496250072116\\
438001	4\\
440001	3\\
442001	2\\
444001	1\\
446001	1\\
448001	1\\
450001	1\\
452001	1\\
454001	1\\
456001	1.58496250072116\\
458001	1.58496250072116\\
460001	1.58496250072116\\
462001	1\\
464001	2\\
466001	2.58496250072116\\
468001	2\\
470001	2\\
472001	2\\
474001	2\\
476001	2.58496250072116\\
478001	2.58496250072116\\
480001	2.58496250072116\\
482001	1.58496250072116\\
484001	1.58496250072116\\
486001	2.58496250072116\\
488001	1\\
490001	1.58496250072116\\
492001	0\\
494001	0\\
496001	0\\
498001	1.58496250072116\\
500001	2.58496250072116\\
502001	1\\
504001	1\\
506001	1\\
508001	1\\
510001	1\\
512001	2\\
514001	2\\
516001	1\\
518001	1\\
520001	0\\
522001	0\\
524001	0\\
526001	0\\
528001	0\\
530001	0\\
532001	0\\
534001	0\\
536001	0\\
538001	0\\
540001	0\\
542001	0\\
544001	1\\
546001	1\\
548001	0\\
550001	0\\
552001	0\\
554001	0\\
556001	0\\
558001	0\\
560001	0\\
562001	0\\
564001	0\\
566001	0\\
568001	0\\
570001	0\\
572001	0\\
574001	0\\
576001	0\\
578001	0\\
580001	0\\
582001	0\\
584001	0\\
586001	1\\
588001	0\\
590001	0\\
592001	0\\
594001	0\\
596001	0\\
598001	0\\
600001	0\\
602001	0\\
604001	0\\
606001	0\\
608001	0\\
610001	0\\
612001	1\\
614001	1\\
616001	0\\
618001	1\\
620001	0\\
622001	1\\
624001	1\\
626001	1\\
628001	1\\
630001	1\\
632001	1\\
634001	0\\
636001	1\\
638001	0\\
640001	1\\
642001	0\\
644001	0\\
646001	0\\
648001	0\\
650001	0\\
652001	0\\
654001	0\\
656001	0\\
658001	0\\
660001	0\\
662001	0\\
664001	0\\
666001	0\\
668001	0\\
670001	0\\
672001	0\\
674001	0\\
676001	0\\
678001	0\\
680001	0\\
682001	0\\
684001	0\\
686001	0\\
688001	0\\
690001	0\\
692001	0\\
694001	0\\
696001	0\\
698001	0\\
700001	0\\
702001	0\\
704001	0\\
706001	0\\
708001	0\\
710001	0\\
712001	0\\
714001	0\\
716001	0\\
718001	0\\
720001	0\\
722001	0\\
724001	0\\
726001	0\\
728001	0\\
730001	1\\
732001	1\\
734001	1\\
736001	1\\
738001	0\\
740001	1\\
742001	1\\
744001	1\\
746001	1\\
748001	1\\
750001	1\\
752001	1\\
754001	1\\
756001	1\\
758001	1\\
760001	1\\
762001	1\\
764001	1\\
766001	1\\
768001	1\\
770001	1\\
772001	1\\
774001	1\\
776001	1\\
778001	1\\
780001	1\\
782001	1\\
784001	1\\
786001	1\\
788001	1\\
790001	1\\
792001	1\\
794001	1\\
796001	1\\
798001	1\\
};
\end{axis}
\end{tikzpicture}%
	\vspace{-0.2cm}
	\caption{Key candidate search space for the 128 bits of the key of the \Sbox\ AES implementation of OpenSSL retrieved using the TSX-based detection and method 2.} \label{flush_count_tot}
	\vspace{-0.5cm}
\end{figure}

Figure~\ref{flush_count_tot} is complementary to Figure~\ref{flush_count}. It represents the number of guesses that an attacker would require to retrieve the key using brute force and the information about the actual access of the victim. Note that both plots represent the same experiment. Different experiments have yielded to similar results but, as we have stated before depend on the actual threshold defined by \textbf{WAIT\_TIME}. Besides, we have noticed that the location of the \Sbox\ in the cache is also important, because there are some sets that are noisier than others. 

\subsection{Attack against RSA}

The modular exponentiation executed for the RSA decryption operations in the wolfSSL implementation is a variation of the well-known square-and-multiply algorithm. Indeed, it is a cache attack protected version of the square-and-multiply implementation that we have described in section \ref{sec:rsa}. 

Based on the code they provide for the tests, we generated different secret keys of 2048 bits, and embedded them in our server application in such a way that it decrypts the received data by calling to the wolfSSL RSA decrypt operation. Note that in this case we can observe the leakage by monitoring accesses to one of the two arrays \lstinline{R[0]} or \lstinline{R[1]}, cause the accesses to each of them depend on the key bit value (0 or 1). During the execution of the multiply function, they are both loaded into the memory, but at the end of the function they perform a copy operation wich only accesses 
the required value. That is, an attacker can for example remove \lstinline{R[0]} from the cache before the execution of the copy operation and check it afterwards. This operation takes around 70-80 cycles in our system, which should be enough for the observation.

However, attacking this implementation is eased by the reduce operation executed after the multiply operation. Such function only loads the correct value of R[y] where y is the secret key value. The execution of the reduce operation lasts about 2300 cycles in our test system. Note that this time even allows the execution of a complete probe cycle so we do not to be so precise evicting the data when targeting this function.

As a difference with the attack against AES where we targeted the \Sbox\ for both detection and eviction it is not possible to use the same content for detection and eviction in this case. Namely, we have used both the multiply and the reduce operations for the detection and later evict R[0]. Note that while the functions are shared, R[0] and R[1] are not, so the approach explained in method 1 is not possible. Also, the attacker will have to profile the application for determining the \textbf{WAIT\_TIME} and to determine the cache set in which R[0] is loaded. The task of profiling is eased with the help of the detection of the multiply function. In our posterior experiments we assume that the attacker already knows the set where R[0] maps.

There are also some other differences with method 2, since we do not require such an accurate eviction and loading the data of a whole eviction set conflicting with R[0] in the transactional region leads to false positives in detection so it is not required. However, some of them can be loaded to reduce the time it takes to evict R[0] considering that all of them can be accessed when retrieving the actual information about the access equivalent to the inference step in algorithm \ref{aessbox}, line \ref{lin:rec}.

In this case, the key bits correctly guessed depend on both the accuracy of the detection and on the ability of the attacker to remove the data from the cache during the execution of the leaky parts of the code. The mean time between the execution of two multiply operations is about 24000 cycles. That time seems to be ``constant'' and it is enough for carrying the detection, eviction and retrieving the data. We collected information for the execution of 100 RSA decryptions. Our attack correctly detected 96.8\% of the multiply operations introducing 1.3\% of false positives. From those correctly detected operations, the information referring to the access to R[0] featured 91\% of true positives rate and 87.2\% of false negative rate, namely a precision of 87,6\%. 

Note that no further processing of the results was done. If so, it is possible to determine which of the executions of the multiply operation were not detected. Note that since we get quite exact timestamps from the aborts, trace alignment becomes easier. The decision about the correct value of the secret bits of the exponent can be made based on the information retrieved from various traces.

\section{Countermeasures}

The presented attack is feasible due to the fact that the interval between an access to a piece of data or a function that could leak data, and the following access gives the attacker the possibility to observe such accesses and, as a consequence, retrieve the secret information. In order to prevent this leakage, these susceptible windows must be removed from the source code, and as a consequence the code should be redesigned. In order to help developers to find leakages in their code, there are tools that detect these leakages \cite{WeiserDATA18,Wichelmann2018}.

In particular, the AES \Sbox\ implementation for x86, could make this attack harder just by including one last prefetch after the encryption ends. Alternatively, each \Sbox\ lookup could access all four cache lines, eliminating cache line leakage.
Considering the wolfSSL RSA implementation, it should at least load into the cache the leaky data in the two vulnerable functions. Note however, that the proposed countermeasures are intended to prevent the exploitation through the LLC, however the powerful SGX scenario may still be able to retrieve some information.

Regarding the countermeasures that prevent cache attacks by means of new cache designs~\cite{180212} or applying hardware modifications \cite{Liu2014,Wang2007}, they can be effective for the presented attack. However, they are not available yet. Similarly, techniques that allocate the victim and the attacker data in different and mutually exclusive cache sets~\cite{7446082_cata} would prevent this attack.

Finally, detection based countermeasures that monitor the execution of the algorithms they aim to protect and collect information about execution times or from performance counters (i.e. cache misses or accesses) to detect changes in the execution trace could detect the attack \cite{ChiappettaEtAl2016,Briongos_2018,Briongos2016}. Note that \attackName\ was not designed to be stealthy and generates cache misses on the victim algorithm. However \attackName\ can also improve the efficiency of existing attacks. Indeed, we tested it against the T-table implementation, retrieving the secret key in about 4-5 ms from just 300 encryptions. These short times seriously limit the capability of the aforementioned countermeasures to trigger the alarm on time. 

\section{Conclusions}
We present \attack , a new side channel attack consisting of the steps \emph{aim}, \emph{wait}, and \emph{shoot}. The \emph{aim} step detects the exact start of the victim operation. The attacker then \emph{waits} for a predetermined time. Last, in the \emph{shoot} step, the target element is evicted from the cache at precisely the right moment. We show that \attack\ can be used to launch last level cache side channel attacks on implementations that are generally regarded as protected by leveraging tiny windows of secret dependent accesses in the implementation. 

Achieving such accurate evictions is possible thanks to the ability of TSX to synchronize with the execution of the target algorithms, the knowledge of the replacement policies implemented in Intel processors and to the fact that the modification of the state of the cache inside the transactional region remains even if the transaction aborts and never commits.

We demonstrate this by retrieving the secret key from the OpenSSL AES \Sbox\ implementation and the secret bits of the modular exponentiation implemented as part of the the wolfSSL RSA decryption algorithm, both of which were regarded as secure against cache side channel attacks since traditional cache attacks would not be able to retrieve such information. \attack\ can however be applied to a much wider range of targets. 

\bibliographystyle{plain}
\bibliography{cache}

\par\leavevmode
\end{document}